\documentclass[aps,prl,reprint,10pt,amssymb,superscriptaddress,showpacs,floatfix]{revtex4-1}
\usepackage{graphicx}
\usepackage{braket}
\usepackage{natbib}
\usepackage{amsmath}

\begin{document}
\title{Can the exciton--polariton regime be defined by its quantum properties?}
\author{D. G. \surname{Su{\'a}rez-Forero}}
\affiliation{Universidad Nacional de Colombia - Bogot\'a, Facultad de Ciencias, Departamento de F{\'isica},
Grupo de {\'O}ptica e Informaci{\'o}n Cu{\'a}ntica, Carrera 30 Calle 45-03, C.P. 111321, Bogot{\'a},
Colombia}

\author{G. \surname{Cipagauta}}
\affiliation{Universidad Nacional de Colombia - Bogot\'a, Facultad de Ciencias,  Departamento de F{\'isica},
Grupo de {\'O}ptica e Informaci{\'o}n Cu{\'a}ntica, Carrera 30 Calle 45-03, C.P. 111321, Bogot{\'a},
Colombia}

\author{H. \surname{Vinck-Posada}}
\email{hvinckp@unal.edu.co}
\affiliation{Universidad Nacional de Colombia - Bogot\'a, Facultad de Ciencias,  Departamento de F{\'isica},
Grupo de {\'O}ptica e Informaci{\'o}n Cu{\'a}ntica, Carrera 30 Calle 45-03, C.P. 111321, Bogot{\'a},
Colombia}

\author{K. M. \surname{Fonseca-Romero}}
\affiliation{Universidad Nacional de Colombia - Bogot\'a, Facultad de Ciencias,  Departamento de F{\'isica},
Grupo de {\'O}ptica e Informaci{\'o}n Cu{\'a}ntica, Carrera 30 Calle 45-03, C.P. 111321, Bogot{\'a},
Colombia}

\author{B. A. \surname{Rodr{\'i}guez}}
\affiliation{Instituto de F{\'i}sica, Universidad de Antioquia, Medell{\'i}n, A.A. 1226, Medell{\'i}n, 
 Colombia}

\begin{abstract}
Using a simple fully quantum model in an effective exciton scheme that takes
into account the system--environment 
interaction, we study the different regimes arising in a microcavity--quantum
dot system. Our numerical calculations of the emission linewidth, emission
energy, integrated intensity and second- and third-order correlation functions
are in good qualitative agreement with reported experimental results. We show
that the transition from the polariton-laser to the photon-laser regime can be
defined through the critical points of both the negativity and the linear
entropy of the steady state. 
\end{abstract}

\maketitle

The solid--state realization of Bose--Einstein Condensation (BEC) has been
achieved in exciton--polariton systems
\cite{Dang,Yamamoto2006PRL97.146402}. These quasi--bosons, arising from the
strong coupling between photons and electron--hole pairs in semiconductor
microcavities ($\mu$-C), have a high critical temperature due to their small
effective mass (eight orders of magnitude smaller than hydrogen atom mass).
After more than two decades of theoretical and experimental investigations,
nowadays it is understood that the large occupation of the polariton ground
state cannot be identified with usual thermodynamic equilibrium BEC states
\cite{Yamamoto,Dang,Snoke}. Instead, the corresponding experimentally observed
regime has been called polariton laser \cite{Dang,Bloch1,Bloch2} because of its
dynamical nature and the gain in the light--emission intensity. The transition
from this regime to a second one, identified with the well-known photon laser,
has also been observed
\cite{Bloch1,Bloch2,Christo}. 
 
Cavity polariton systems have been studied from different theoretical
perspectives. Assuming thermal equilibrium, a trial wave function that takes
into account the coherence properties of both light and matter, has been able to
predict multiple phase transitions \cite{Littlewood,Tetsuo}. On the other hand,
when the matter--light state is obtained from some equation of motion (mean
field dynamics \cite{VinckPosada,Savona,Kavokin}, master equation in
multiexcitonic scheme \cite{Gonzalez}, dissipative Jaynes--Cummings model
\cite{Tejedor,Fabrice1,Fabrice2,Fabrice3,Vera,Elena,Gerard,Finley}), the
dynamical character of the polariton laser regime is conspicuous, and the
non-Gibbsian character of the stationary state of the system is revealed.
However, the elucidation of the mechanisms behind the appearance of the
different observed regimes is still an open problem. Our work is a first step in
this direction.

The aim of this Letter, the identification of the regimes observed in current
experiments from the quantum properties of the steady state, is possible due to
the simplicity of our model. Indeed, we consider a single pumped radiator
interacting with a leaky mode of the electromagnetic field, ignoring collective
effects. Our calculations, however, are in qualitative good agreement with the
experimental results. Moreover, we can correlate the entanglement, mixedness and
the coherence functions of the steady state not only with  the observed regimes
but also with important physical parameters like the pumping rate
and the detuning. In addition, we are able to provide a criterion to identify
the ``best'' polariton that can be sustained by the system. 

We model a  quantum dot (QD)  embedded in a $\mu$-C, as a two-level system (ground $\ket{G}$ and excited $\ket{X}$ states).
Its interaction with a single electromagnetic mode of frequency $\omega_C$, in the dipole and rotating wave approximations, is described by the Hamiltonian ($\hbar=1$):
\begin{equation}
\hat{H}=\omega_C\hat{a}^{\dagger}\hat{a}+(\omega_C-\Delta)\hat{\sigma}^{\dagger}\hat{\sigma}+g(\hat{a}\hat{
\sigma }^{\dagger}+\hat{a}^{\dagger}\hat{\sigma}).
\label{eq:1}
\end{equation}
The detuning $\Delta$ is the difference between cavity mode and exciton
energies, $g$ is the matter--light coupling constant,
$\hat{\sigma}=\ket{G}\bra{X}$ is the QD ladder operator, and $\hat{a}^{\dagger}$
($\hat{a}$) is the usual creation (annihilation)
operator of the cavity mode. The Hamiltonian (\ref{eq:1}) commutes with the excitation number
$\hat{N}=\hat{N}_{ph}+\hat{N}_{ex}=\hat{a}^{\dagger}\hat{a}+\hat{\sigma}^{
\dagger}\hat{\sigma}$; hence, it only causes transitions between matter--light
states of the same excitation manifold. Polaritons are defined as the energy
eigenstates $\hat{H}$, and are explicitly given by
\begin{eqnarray}
\ket{n,+}&=&\sin{\Phi_n}\ket{G,n}+\cos{\Phi_n}\ket{X,n-1} \nonumber\\
\ket{n,-}&=&\cos{\Phi_n}\ket{G,n}-\sin{\Phi_n}\ket{X,n-1},
\label{eq:2}
\end{eqnarray}
where $\{\ket{n}\}$ denotes the Fock number states of the field and $\tan{2\Phi_n}=2g\sqrt{n}/\Delta$. 
We include two non-conservative processes, the loss of photons in the $\mu$-C
($\kappa$) and the continuous pumping of excitons ($P$), in the master equation
for the density operator $\hat{\rho}$ of the system 
\begin{eqnarray}
\frac{d\hat{\rho}}{dt} &=& i[\hat{\rho},\hat{H}]+
\tfrac{1}{2}P(2\hat{\sigma}^{\dagger}\hat{\rho}\hat{\sigma}-\hat{\sigma}\hat{\sigma}^{\dagger}\hat
{ \rho } -
\hat{\rho}\hat{\sigma}\hat{\sigma}^{\dagger}) \nonumber \\
&+&\tfrac{1}{2}\kappa(2\hat{a}\hat{\rho}\hat{a}^{\dagger}-\hat{a}^{\dagger}\hat{a}\hat{\rho}-\hat{\rho}\hat{a}
^ {\dagger}\hat{a}),
\label{eq:3}
\end{eqnarray}
where we have made the Born-Markov approximation. We neglect other
system--environment interaction mechanisms (e.g., spontaneous emission,
dephasing, photon pumping, polariton pumping, etc.) because their effect is
either small or is already effectively contained in the master equation. If a
better adjustment with the experimental results is desired those mechanisms can
be included and fitted \cite{Finley}, but the qualitative physical image remains
essentially unchanged.

The basic assumption behind our approach, which focuses on the steady state
$\hat{\rho}_{\textrm{ss}}$ of the equation of motion (\ref{eq:3}), is that
polariton lifetime is much longer than the time required to reach the asymptotic
solution \cite{Yamamoto2006PRL97.146402}. The steady state
$\hat{\rho}_{\textrm{ss}} =\hat{\rho}(\kappa,P,g,\Delta)$ of the system, is a 
function of the dissipative rates $\kappa$ and $P$, the matter-light coupling
constant $g$ and the detuning $\Delta$. In the weak-coupling regime $g\ll P,
\kappa$, the pumping keeps the QD in its excited state while the dissipation
steers the electromagnetic field to its ground state. Other states are not
significantly populated because matter excitation cannot be converted into
photons. On the other extreme, the ultra-strong coupling regime $g\ggg P,
\kappa, \Delta$, the long-time density matrix becomes (almost) diagonal in the
basis of bare states $\ket{G/X,n}$. The larger the coupling $g$ the smaller the
difference of the populations of $\ket{G,n}$ and $\ket{X,n-1}$ ($\propto
1/g^2$). The coherences $|\rho_{GnXn-1}|$, which decay
as $1/g$, also vanish as the coupling $g$ increases. 

In this work we focus on the strong-coupling regime $g\gg P, \kappa$, in which
the coherences $\rho_{GnXn-1}$ are small, but different from zero. In resonance
they are purely imaginary. For small detunings they acquire a small real part.
If the detuning increases, $\Delta\gg g$, the matter-light interaction becomes
dispersive, i.e., the energies of the matter states depend on the number of
photons, and the mechanism which converts matter excitations into photons is
suppressed. Thereby, large detunings correspond to a weak coupling regime. We
conclude that in the regime $|\Delta|\sim g\gg P, \kappa$, the steady state of
the system is expected to exhibit a polaritonic behavior. Unless stated
otherwise, the steady state solution $\hat{\rho}_{ss}$ of (\ref{eq:3}) is
obtained for the initial condition $\hat{\rho}(0)=\ket{G0}\bra{G0}$, and
setting $\omega_C=1$ eV, $g=1$ meV and $\kappa=5\times10^{-2}$ meV, while
$\Delta$ and $P$ are varied in ranges similar to those of current experiments
\cite{Bloch1,Bloch2,Reithmaier2004Nature432.197}. 

Evidence of the spontaneous coherence build-up associated with polariton states
are currently detected through the photoluminescence properties in quantum wells
(QW) \cite{Bloch1,Bloch2,Dang}. We compare our theoretical predictions in QDs
with the experimental findings in QWs because: $i$) due to experimental
difficulties no analogous results for QDs have been reported and $ii$) it is
reasonable to assume that some of the physical mechanisms behind the
exciton--polariton laser regime are the same in both cases. Additionally, the
present approach may shed light on the separation of the collective effects in
QWs from those of a QD single emitter. 

\begin{figure}
\includegraphics[width=0.99\columnwidth]{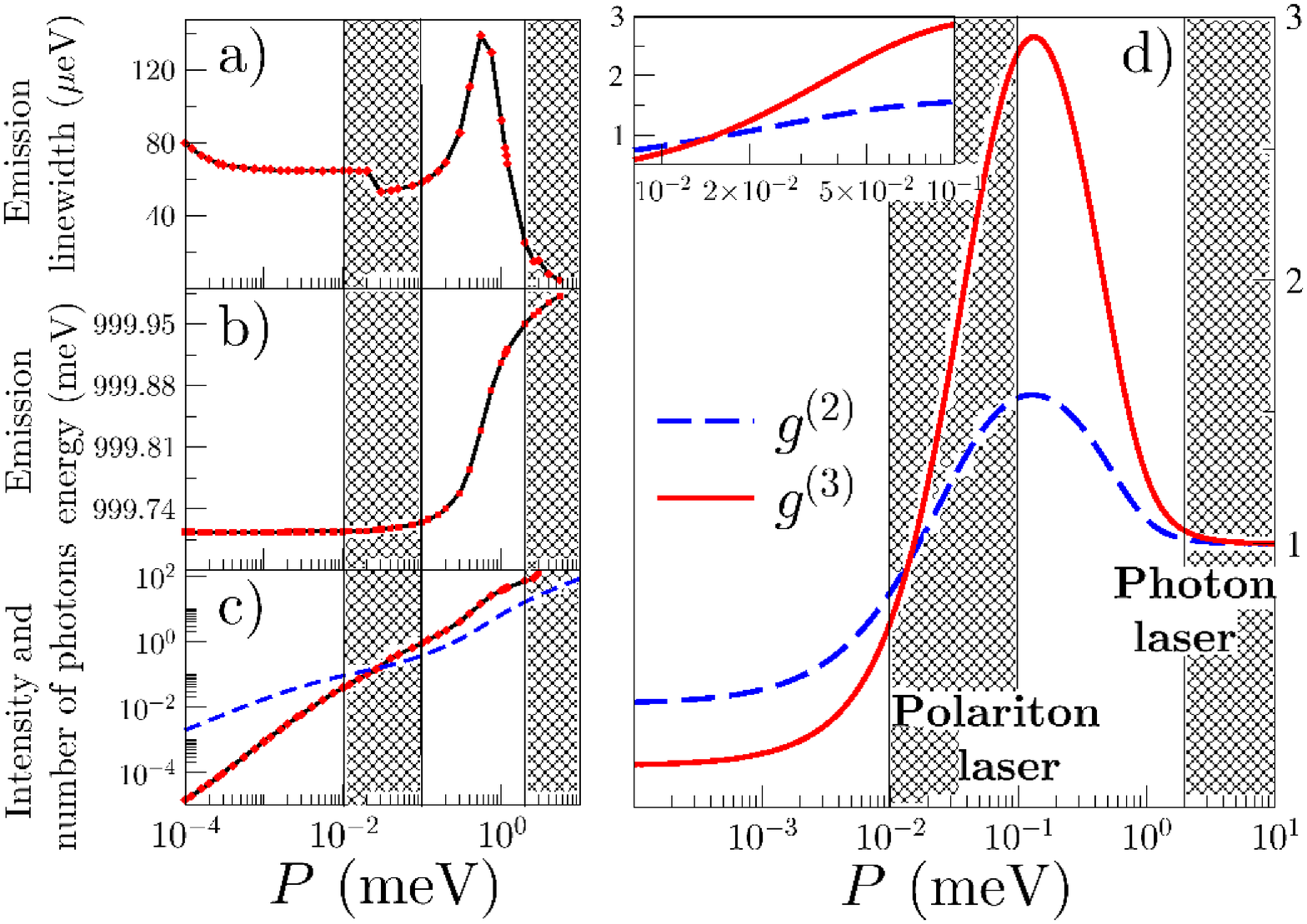}
\caption{\label{fig1}(color online). (a) Emission linewidth, (b) emission energy, (c) integrated intensity (continuous line) and
average number of photons (dashed line) and (d) second-- and third--order correlation functions versus the incoherent exciton pumping $P$, for $\kappa=5\times10^{-2}$ meV and $\Delta=2.5$ meV. The marked regions correspond to the polariton--laser and photon--laser regimes.}
\end{figure}

The calculated emission linewidth, emission energy, integrated intensity and
number of photons are shown in fig. \ref{fig1} as a function of the pumping
rate. This numerical calculation used the quantum regression theorem
\cite{Tejedor,Fabrice1,Fabrice2,Vera} and the integrated intensity
corresponds to the area under the curve of the peak associated to the
transition between the polaritons in the manifolds $n=1$ and $n=2$. The
linewidth (fig.\ref{fig1}.a) exhibits the characteristic reduction in the
polariton regime, and the subsequent growth and decrease
\cite{Bloch1,Bloch2,Dang}. The emission energy blueshifts (fig.\ref{fig1}.b)
from the exciton transition frequency to the cavity mode frequency. In the
polaritonic region the blueshift is smaller than observed in QWs
\cite{Bloch1,Bloch2}. In the intermediate region, $10^{-1}$ meV $\lesssim
P\lesssim1$ meV, the calculated blueshift grows faster than the measured one.
Despite the small slope changes in the integrated intensity as a function of $P$
(fig.\ref{fig1}.c) our results are consistent with the previous prediction of
absence of threshold in a one--atom laser \cite{Carmichael1994PRA50.4318}.
However, the nonlinearity of the model is evident in the curve of the average
number of photons. Only for large values of the pumping ($P\gtrsim1$ meV), this
curve is parallel to that of the integrated emission intensity (whose slope is
approximately one).

Statistics of the emitted light can be characterized by the normalized
second-- $g^{(2)}(0)=\braket{\hat{a}^\dag \hat{a}^\dag
\hat{a}\hat{a}}/\braket{\hat{a}^\dag \hat{a}}^{_2}$ and third--order
$g^{(3)}(0)=\braket{\hat{a}^\dag \hat{a}^\dag \hat{a}^\dag
\hat{a}\hat{a}\hat{a}}/\braket{\hat{a}^\dag \hat{a}}^{_3}$ coherence functions,
plotted in fig.(\ref{fig1}.d). For small pumping power
($P\lesssim2\times10^{-2}$ meV) $g^{(2)}(0),g^{(3)}(0)<1$, a footprint of
quantum-like light, i.e. the partial state of the field is nearly a Fock state
with small number of photons, as expected. Indeed, our calculations show that in
this regime
$\hat\rho_{ss}\approx\ket{G}\bra{G}\otimes\{\sin{\varphi}\ket{0}\bra{0}+\cos{
\varphi}\ket{1}\bra{1}\}$, with $\varphi\approx 0$. For intermediate values of
the pumping ($10^{-2}$ meV $\lesssim P\lesssim10^{-1}$ meV) both correlations
functions monotonically grow beyond one (inset fig.1.d), as has been
experimentally observed \cite{Horikiri2010PRB81.033307}. We analyze this
behavior below in the text. For larger pumping rates ($P\gtrsim 2$ meV), the
state of the field becomes coherent up to third order (when
$g^{(2)}(0)=g^{(3)}(0)=1$), and thus the linewidth falls (fig.1.a). In this
region, the state of the field obtained by the partial trace over the excitonic
degrees of freedom, has a fidelity of more than 0.99 with the random-phase
coherent state
$\int_0^{2\pi}\tfrac{d\phi}{2\pi}\Ket{\alpha e^{i\phi}}\Bra{ \alpha e^{i\phi}}$,
where $\Ket{\alpha e^{i\phi}}$ is an usual coherent state with
$|\alpha|^{2}$ average number of photons. This type of state has been proposed
to describe the features of a (true) photon-laser \cite{Molmer1997PRA55.3195}.

\begin{figure}
\includegraphics[width=0.99\columnwidth]{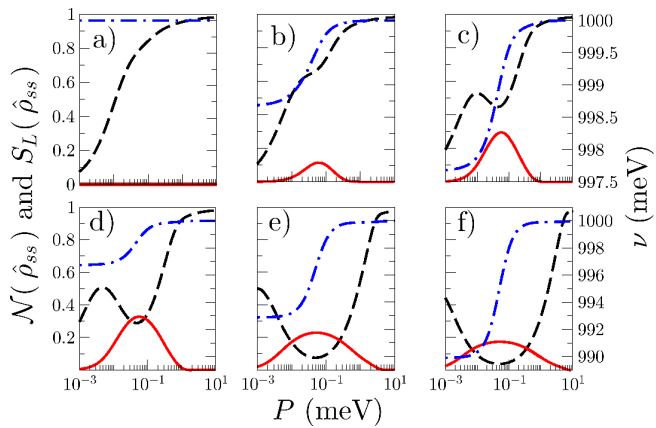}
\caption{\label{fig2} (color online). $\mathcal{N}(\hat{\rho}_{ss})$ (continuous red line), $S_{L}(\hat{\rho}_{ss})$ (dashed black line) and $\nu$ (dashed-dotted blue line) as a function of the incoherent exciton pumping $P$, for $\kappa=5\times10^{-2}$ meV and (a) $\Delta=0$, (b) $\Delta=1$ meV, (c) $\Delta=2$ meV, (d) $\Delta=3$ meV, (e) $\Delta=7$ meV and (f) $\Delta=10$ meV. Note that for $\Delta\neq0$ the maximum of $\mathcal{N}(\hat{\rho}_{ss})$, the minimum of $S_{L}(\hat{\rho}_{ss})$ and the inflection point of $\nu$ coincide.}
\end{figure}

We stress that the regimes identified through the characteristics of the emitted
light, mirror quantum properties of the system state. We describe those quantum
properties by entanglement and mixedness of the steady state. Linear entropy and
negativity are employed to quantify mixedness and matter-light entanglement,
respectively. The former, defined as
$S_L(\hat{\rho})=1-\textrm{Tr}\hat{\rho}^{_2}$, vanishes for pure states and is
maximum for maximally mixed states. The latter is defined as
$\mathcal{N}(\hat{\rho})=2\sum_{\lambda<0}|\lambda|$, where $\lambda$ denotes
the eigenvalues of the partial transpose of $\hat{\rho}$
\cite{Zyczkowski1998PRA58.883,VidalPRA65.032314}. Finally, searching for a
relation between the energy of the system and its quantum characteristics (such
as entanglement), we introduce the differential energy per excitation,
$\nu(\kappa,P)=\tfrac{1}{2} d\braket{\hat{H}}/d\braket{\hat{N}}$, as a
convenient measure of energy per particle. The factor of 2 in the definition of
$\nu$ has been  chosen to satisfy the condition $\nu(P\ll g,\kappa\ll
\Delta)\approx \omega_C-\Delta$. The negativity, linear entropy and the
differential energy per excitation are depicted in fig. \ref{fig2}. 

Assuming the strong-coupling regime, the formation of the polariton is hindered
by two different mechanisms, which depend on the detuning. While the mixedness
of the state is large for small detuning (fig. \ref{fig2}.a and \ref{fig2}.b),
matter and light decouple for large detuning (fig. \ref{fig2}.f). In the
intermediate region $g\lesssim |\Delta|\lesssim 10 g$, where the polariton is
well defined, both mechanisms compete. The matter-light entanglement is
enhanced, and the entropy of the asymptotic state of the system decreases with
increasing detuning. The negativity of the steady state
$\mathcal{N}(\hat{\rho}_{ss})$ vanishes for small and large values of the
pumping power, and attains a maximum at the point $P=\kappa$ --the mid-point of
the polariton-laser region. 

Now, we are able to give a possible explanation of the behavior of the
correlation functions in the polariton regime. The asymptotic state in the
polaritonic region is a mixed entangled state which satisfies 
$g^{(3)}(0)>g^{(2)}(0)>1$. This behavior of the correlation functions is a
rather generic feature. As an example, we consider another mixed entangled state
$\hat\rho_{pol}(\bar{n}) =\sum_n P_n(\bar{n})\ket{n,+}\bra{n,+}$. Since the
probabilities  $P_n(\bar{n})=e^{-\bar{n}} \bar{n}^n/(n!)$ are Poisson weights,
this is a polariton coherent state. However, the reduced photon state has
super-poissonian statistics, i.e., the second- and third-order coherence
functions can not be expected to be unity. Hence, matter-light entanglement is a
viable alternative to the standard explanation of the unexpected behavior of
these functions, based on polariton-polariton and polariton-phonon
interactions. 

\begin{figure}
\includegraphics[width=0.98\columnwidth]{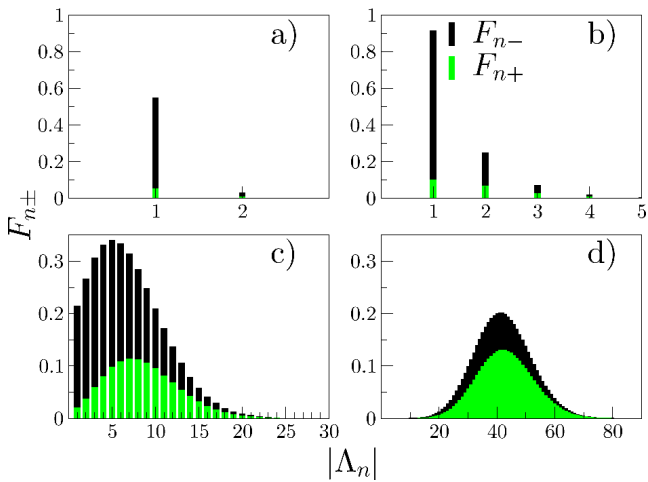}
\caption{\label{fig4}(color online). Sequence of non-zero fidelities $F_{n\pm}$ between the steady state $\hat{\rho}_{ss}$ and the $\Lambda_n$--lower(black)$/$upper(green) polaritons $\ket{n,\pm}$ for $\kappa=5\times10^{-2}$ meV, $\Delta=3$ meV and (a) $P=2\times10^{-3}$ meV, (b) $P=5\times10^{-2}$ meV, (c) $P=1$ meV and (d) $P=5$ meV. $|\Lambda_n|$ denotes the excitation number of the polariton manifold $\Lambda_n$.}
\end{figure}

The differential energy per excitation $\nu$ is plotted in fig. \ref{fig2} as a
function of $P$ for several values of the detuning. For $\Delta=0$, $\nu$ is
independent of $P$ and equals to the cavity mode energy, since the number of
photons becomes much larger than the number of matter excitations. Our numerical
results show that for $|\Delta|\gtrsim g$, and small ($P<10^{-2}$ meV) or large
($10$ meV $>P>10^{-1}$ meV) pumping rates, $\nu$ is almost a constant, equal to
the exciton (photon) energy in the former (latter) case. The same constant
values would had been obtained with the definition $\tilde{\nu}=H/N$. For
intermediate values of $P$, $\nu$ displays an inflexion point at $P=\kappa$,
where $\nu=\omega_C-(\Delta/2)$ is halfway between the exciton and photon
energies.  Moreover, since the mean number of excitations is one, it is tempting
to define $P=\kappa$ as the condition for the ``optimum'' polariton. In order to
quantify this idea we compare the steady state with the polariton states
$\ket{n,\pm}$ defined by \eqref{eq:2}, using the sequence of non-zero fidelities
$F_{n\pm}=\sqrt{\braket{n,\pm|\hat{\rho}_{ss}|n,\pm}}$. For small values of $P$
(fig.\ref{fig4}.a), when $\rho_{G0G0}$ is much larger than the other
populations, only $F_{1-}$ does not vanish --however it is relatively small--.
For $P=\kappa$ (fig.\ref{fig4}.b) $F_{1-}$ increases up to more than 0.95, while
the remaining fidelities are still small. Hence, the steady state
$\hat{\rho}_{ss}$ is quite similar to the $\Lambda_1$--polariton $\ket{1-}$. As
$P$ increases, $F_{n\pm}$ is non-zero for larger excitation-numbers, but their
values are very small.

The  excitation number ($\hat{N}$) symmetry associated with the Hamiltonian
(\ref{eq:1}) is broken in the time evolution provided by the master equation
(\ref{eq:3}), in the sense that the asymptotic state of the system cannot be
labeled with a single eigenvalue of $\hat{N}$. If polariton-like behavior is
actually present, restoration of the symmetry is expected. In order to account
for this effect, we introduce the participation ratio $PR=\sum_{n=0}^\infty
P_n^2$, where $P_n$ is the probability to have $n$ excitations in the asymptotic
state. This quantity, which varies from zero --all excitation numbers are
equiprobable-- to one --only one occupied manifold--, displays a global maximum
at zero pumping rate and a local maximum at $P\approx\kappa$, in the strong
coupling regime, signaling a partial restoration of symmetry. This can be
understood as a combined effect of the decrease of the mixedness of the state
and the increase of its entanglement, occurring at $P\approx\kappa$, as discussed
above (again, $\Delta\sim g$).

\begin{figure}
\includegraphics[width=0.90\columnwidth]{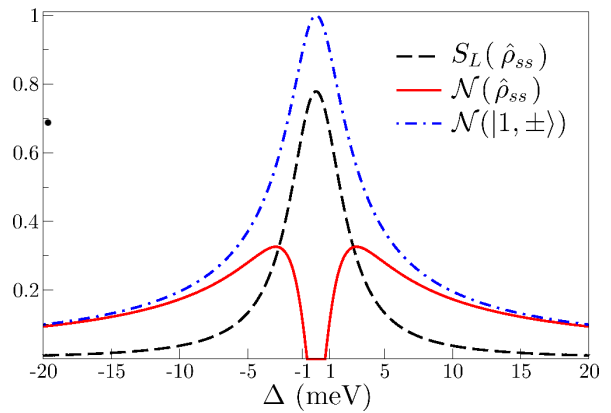}
\caption{\label{fig3}(color online). Linear entropy (dashed black line) and negativity (continuous red line) of the steady state of the system $\hat{\rho}_{ss}$, and negativity of polaritons of the excitation manifold $\Lambda_1$ (dashed-dotted blue line) as a function of $\Delta$ for $P=\kappa=5\times10^{-2}$ meV.}
\end{figure}

As we have seen, the optimum polariton exhibits maximum negativity, minimum
linear entropy, a local maximum of the participation ratio and an inflection
point of the differential energy per excitation, provided that the system is in
strong coupling. To quantify our previous qualitative arguments --which show
that polaritons cannot be sustained neither for small nor for large detunings--,
it is worth examining the behavior of the linear entropy and the negativity at
$P=\kappa$, as a function of the detuning (fig.\ref{fig3}). Three regions can be
identified. In the first region, $|\Delta|< 0.68$ meV, the negativity of the
steady state is exactly zero and the linear entropy is larger than $0.72$. In
the second region, $ 0.68$ meV $<|\Delta|\lesssim 10$ meV, the linear entropy
and the negativity are still significative. In the third region the dissipative
polariton is very close to the Hamiltonian polariton $\ket{1,\pm}$, for the
corresponding $\Delta$. Nevertheless, the dressed states (\ref{eq:2}) are
virtually the bare states.

Our results might provide a guide to experiments, in the sense that, in the
strong-coupling regime, both the pumping rate and the detuning have to be
carefully adjusted. In our model, the best matter-light correlation properties
occur at $P=\kappa$ and $\Delta\approx 3g$, where $\mathcal{N}\approx0.32$ and
$S_L\approx0.29$.  From the theoretical point of view we propose the following
criterion: if the negativity rises above 0.25, close to its maximum possible
value, its inflection points, as a function of the pumping power, can be used to
define the polaritonic regime. When this condition is fulfilled all the
quantifiers that we have examined exhibit a characteristic change. The emission
energy presents a blueshift, the differential energy per excitation has an
inflection point, the emission line decreases, the second- and third-order
correlation functions increase beyond their value for photon coherent states, the
entropy decreases and the negativity increases. With the exception of the first
two, these changes can be understood as a coherence gain of the asymptotic state
of the system. 

We are grateful with Prof. P.S.S. Guimar{\~a}es from UFMG (Brazil), Prof. J.
Mahecha from UdeA (Colombia) and Dra. J. Restrepo from UAN (Colombia) for
critical reading of the manuscript. The authors acknowledge partial financial
support from Direcci\'on de Investigaci\'on - Sede Bogot\'a, Universidad
Nacional de Colombia (DIB-UNAL) under project 12584, and technical and
computational support from Grupo de \'Optica e Informaci\'on Cu\'antica
(GOIC-UNAL), and Grupo de F\'isica At\'omica y Molecular (GFAM-UDEA).

\bibliography{BIB} 
\end{document}